\documentclass{article}
\usepackage{spconf,amsmath,graphicx}
\usepackage[utf8]{inputenc}
\usepackage{amsfonts}
\usepackage{amssymb}
\usepackage[ruled,vlined, linesnumbered]{algorithm2e}
\usepackage{algorithmicx}
\usepackage[noend]{algpseudocode}
\usepackage[dvipsnames]{xcolor}
\usepackage{booktabs}
\usepackage{multirow}

\title{SPEECH MODELING WITH A HIERARCHICAL TRANSFORMER DYNAMICAL VAE}
\name{Xiaoyu Lin$^1$, Xiaoyu Bie$^1$, Simon Leglaive$^2$, Laurent Girin$^3$, Xavier Alameda-Pineda$^1$ \thanks{This research was supported by ANR-3IA MIAI (ANR-19-P3IA-0003), ANR-JCJC ML3RI (ANR-19-CE33-0008-01), H2020 SPRING (funded by EC under GA \#871245).}}
\address{$^1$ Inria Grenoble Rh\^one-Alpes, Univ. Grenoble Alpes, France \\
$^2$ CentraleSupélec, IETR (UMR CNRS 6164), France\\
$^3$ Univ. Grenoble Alpes, CNRS, Grenoble-INP, GIPSA-lab, France
}
\begin{document}
\ninept
\maketitle
\begin{abstract}
The dynamical variational autoencoders (DVAEs) are a family of latent-variable deep generative models that extends the VAE to model a sequence of observed data and a corresponding sequence of latent vectors. In almost all the DVAEs of the literature, the 
temporal dependencies within each sequence and across the two sequences are modeled with recurrent neural networks. In this paper, we propose to model speech signals with the Hierarchical Transformer DVAE (HiT-DVAE), which is a DVAE with two levels of latent variable (sequence-wise and frame-wise) and in which the temporal dependencies are implemented with the Transformer architecture. We show that HiT-DVAE outperforms several other DVAEs for speech spectrogram modeling, while enabling a simpler training procedure, revealing its high potential for downstream low-level speech processing tasks such as speech enhancement.
\end{abstract}
\section{Introduction}
\label{sec:intro}

Dynamical variational autoencoders (DVAEs) are a class of latent-variable deep generative models (LV-DGM) that extends the widely-used variational autoencoder (VAE) \cite{Kingma2014, pmlr-v32-rezende14} to model correlated sequences of data \cite{MAL-089}. DVAEs consider a sequence of high-dimensional data $\mathbf{s}_{1:T}$ and a corresponding sequence of low-dimensional latent vectors $\mathbf{z}_{1:T}$, and model the time dependencies within $\mathbf{s}_{1:T}$, within $\mathbf{z}_{1:T}$, and across $\mathbf{s}_{1:T}$ and $\mathbf{z}_{1:T}$, using deep neural networks. In all the DVAE models reviewed in \cite{MAL-089}, e.g., \cite{https://doi.org/10.48550/arxiv.1511.05121, NIPS2015_b618c321, 10.5555/3157096.3157343, 10.5555/3294771.3294950}, the 
temporal dependencies are implemented with recurrent neural networks (RNNs). 

DVAEs have achieved great success in modeling sequential data. In particular, they have been successfully applied to the modeling of speech signals, in analysis-resynthesis tasks \cite{NIPS2015_b618c321,10.5555/3294771.3294950, Bie2021ABO} or in speech enhancement  \cite{9053164, 9894060}. Meantime, the use of RNNs in DVAEs also brings some related problems. A classical issue is the mismatch between training and test conditions. 
Indeed, the commonly adopted configuration for RNN training is to use the ground-truth past observed vectors $\mathbf{s}_{1:t-1}$ in the generative model, a training strategy often referred to as \emph{teacher-forcing} \cite{6795228}. However, at test/generation time, we can only use the previously generated values $\hat{\mathbf{s}}_{1:t-1}$ to generate the current one. This generally results in large accumulated prediction errors along the sequence. Directly training an RNN in the generation mode is also difficult. To remedy this problem, \emph{scheduled sampling} can be adopted, i.e., the ground truth past vectors $\mathbf{s}_{1:t-1}$ are progressively replaced with the previously generated ones $\hat{\mathbf{s}}_{1:t-1}$ along the training iterations \cite{10.5555/2969239.2969370}. This strategy was successfully adopted for the training of autoregressive (AR) DVAEs, i.e. DVAEs with a recursive structure on $\mathbf{s}$ \cite{MAL-089}. However, this requires a well designed sampling scheduler to guarantee the prediction performance. 
Besides, RNNs are poorly suited for parallel computation and face gradients exploding and vanishing for long sequences.

Replacing RNNs with Transformers \cite{10.5555/3295222.3295349} for sequential modeling has become a new trend in both natural language processing, e.g., \cite{Devlin2019BERTPO},
and computer vision, e.g.,  \cite{https://doi.org/10.48550/arxiv.2010.11929}. Compared to the recursive computing in RNNs, the Transformer architecture uses attention mechanisms to learn global dependencies between input and output, which allows modeling complex time correlations. Thus, in principle, replacing RNNs with Transformers in DVAEs can also benefit to the modeling of temporal dependencies between the observed and latent sequences. Following this idea, the Hierachical Transformer DVAE (HiT-DVAE) model proposed in \cite{https://doi.org/10.48550/arxiv.2204.01565} combines a DVAE with Transformers for 3D Human motion generation. In this paper, (a) we propose an improved version of HiT-DVAE, named LigHT-DVAE, and (b) we apply it to speech signal modeling. The proposed LigHT-DVAE model reduces the number of parameters by sharing the parameters of the decoders of the original HiT-DVAE model. Experimental results show that we achieve very competitive results in the speech analysis-resynthesis task. We also investigate the modeling capacity of LigHT-DVAE compared to other DVAEs through extensive ablation studies on the model structure. In particular, we show that the structure of LigHT-DVAE makes it robust to using teacher-forcing at training time, which largely simplifies the training procedure compared with other AR DVAEs. To the best of our knowledge, this is the first AR DVAE model that can be trained in teacher-forcing \emph{and} generalizes well to the generation mode. 
\section{The LigHT-DVAE for speech modeling}
\label{sec:HiT-DVAE_model}

\begin{figure*}[htb]
    \centering
    \includegraphics[width=.95\linewidth]{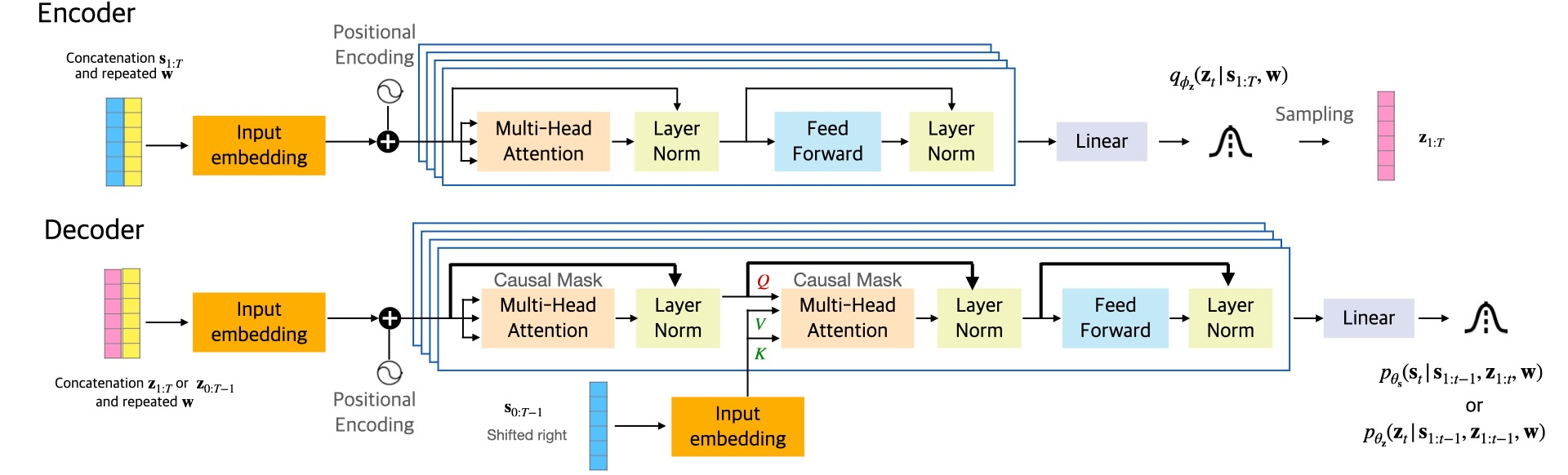}
    \caption{LigHT-DVAE model structure.}
    \label{fig:model_structure}
\end{figure*}

The LigHT-DVAE model is a DVAE in which the temporal dependencies between and within the observed and latent sequences are implemented with Transformers instead of RNNs \cite{https://doi.org/10.48550/arxiv.2204.01565}. In this section, we present LigHT-DVAE for speech signal modeling. We work with the short-time Fourier transform (STFT) of speech waveforms, denoted by $\mathbf{s}_{1:T} \in \mathbb{C}^{F \times T}$. Each vector $\mathbf{s}_t=\{s_{f,t}\}_{f=1}^F$ of the sequence is the short-time complex-valued spectrum at time frame $t$, and $f$ denotes the frequency bin. 

\subsection{Generative model} 
\label{subsec:generative_model}

In addition to $\mathbf{s}_{1:T}$ and the associated sequence of latent vectors $\mathbf{z}_{1:T} \in \mathbb{R}^{L_z \times T}$, with $L_z \ll F$, LigHT-DVAE also defines a time-independent latent variable $\mathbf{w} \in \mathbb{R}^{L_w}$, which is designed to embed high-level features at the speech occurrence level (e.g., speaker ID). The generative model of LigHT-DVAE is defined as:
\begin{align} \label{generative-joint-distribution-hit-dvae}
     & p_{\theta}(\mathbf{s}_{1:T}, \mathbf{z}_{1:T}, \mathbf{w}) = p_{\theta_{\mathbf{w}}}(\mathbf{w}) \nonumber \\ 
     & \ \  \times \prod\nolimits_{t=1}^T p_{\theta_{\mathbf{s}}}(\mathbf{s}_t|\mathbf{s}_{1:t-1}, \mathbf{z}_{1:t}, \mathbf{w})p_{\theta_{\mathbf{z}}}(\mathbf{z}_t|\mathbf{s}_{1:t-1}, \mathbf{z}_{1:t-1}, \mathbf{w}),
\end{align}
where $\theta = \theta_{\mathbf{s}} \cup \theta_{\mathbf{z}} \cup \theta_{\mathbf{w}}$.
As usually adopted in statistical audio/speech processing \cite{6797100, 5720325}, the STFT coefficient at each time-frequency bin $s_{t,f}$ is assumed to follow a circularly-symmetric zero-mean complex Gaussian distribution.\footnote{that we denote $\mathcal{N}_c\big(\cdot; \boldsymbol{\mu}, \boldsymbol{\Sigma})$, with $\boldsymbol{\mu}$ and $\boldsymbol{\Sigma}$ being the mean vector and  covariance matrix, respectively.} The STFT coefficients at different frequency bins are assumed to be independent. The conditional distribution of $\mathbf{s}_t$ in \eqref{generative-joint-distribution-hit-dvae} thus writes:
\begin{equation}
    p_{\theta_{\mathbf{s}}}(\mathbf{s}_t|\mathbf{s}_{1:t-1}, \mathbf{z}_{1:t}, \mathbf{w}) = \mathcal{N}_c\big(\mathbf{s}_t; \mathbf{0}, \text{diag}(\mathbf{v}_{\theta_{\mathbf{s}},t})\big).
    \label{eq:gen-s}
\end{equation}
 In this work, all covariance matrices are assumed diagonal and are represented by the vector of diagonal entries. Here, $\mathbf{v}_{\theta_{\mathbf{s}},t} \in \mathbb{R}_+^F$ depends on the variables $\{\mathbf{s}_{1:t-1}, \mathbf{z}_{1:t}, \mathbf{w}\}$. Note that in practice, HiT-DVAE will input speech power spectrogram, i.e., the squared modulus of $\mathbf{s}_{1:t-1}$, instead of the complex-valued STFT coefficients themselves. 
The latent vector $\mathbf{z}_t$ is assumed to follow a (real-valued) Gaussian distribution: 
\begin{equation}
    p_{\theta_{\mathbf{z}}}(\mathbf{z}_t|\mathbf{s}_{1:t-1}, \mathbf{z}_{1:t-1}, \mathbf{w}) = \mathcal{N}\big(\mathbf{z}_t; \boldsymbol{\mu}_{\theta_{\mathbf{z}},t}, \text{diag}(\mathbf{v}_{\theta_{\mathbf{z}}, t})\big).
    \label{eq:gen-z}
\end{equation}
The mean and variance vectors $\boldsymbol{\mu}_{\theta_{\mathbf{z}},t} \in \mathbb{R}^{L_z}$ and $\mathbf{v}_{\theta_{\mathbf{z}}, t} \in \mathbb{R}^{L_z}_{+}$ both depend on $\{\mathbf{s}_{1:t-1}, \mathbf{z}_{1:t-1}, \mathbf{w}\}$. As for the latent vector $\mathbf{w}$, it is assumed to follow a standard Gaussian prior: $p_{\theta_{\mathbf{w}}}(\mathbf{w}) = \mathcal{N}(\mathbf{w}; \mathbf{0}, \mathbf{I})$.

\subsection{Inference Model}

LigHT-DVAE inference model approximates the intractable exact posterior distribution of the latent sequence and is defined as:
\begin{equation}
    q_{\phi}(\mathbf{z}_{1:T}, \mathbf{w}|\mathbf{s}_{1:T}) = q_{\phi_{\mathbf{w}}}(\mathbf{w}|\mathbf{s}_{1:T}) \prod\nolimits_{t=1}^T q_{\phi_{\mathbf{z}}}(\mathbf{z}_{t}|\mathbf{s}_{1:T}, \mathbf{w}),
\end{equation}
with $\phi = \phi_{\mathbf{w}} \cup \phi_{\mathbf{z}}$,
\begin{equation}
    q_{\phi_{\mathbf{w}}}(\mathbf{w}|\mathbf{s}_{1:T}) = \mathcal{N}\big(\mathbf{w}; \boldsymbol{\mu}_{\phi_{\mathbf{w}}}, \text{diag}(\mathbf{v}_{\phi_{\mathbf{w}}})\big),
    \label{eq:inf-w}
\end{equation}
\begin{equation}
    q_{\phi_{\mathbf{z}}}(\mathbf{z}_t|\mathbf{s}_{1:T}, \mathbf{w}) = \mathcal{N}\big(\mathbf{z}_t; \boldsymbol{\mu}_{\phi_{\mathbf{z}},t}, \text{diag}(\mathbf{v}_{\phi_{\mathbf{z}}, t})\big), \label{eq:inf-z}
\end{equation}
and where $\boldsymbol{\mu}_{\phi_{\mathbf{w}}} \in \mathbb{R}^{L_w}$ and $\mathbf{v}_{\phi_{\mathbf{w}}} \in \mathbb{R}^{L_w}_{+}$ depend on $\mathbf{s}_{1:T}$, and  $\boldsymbol{\mu}_{\phi_{\mathbf{z}},t} \in \mathbb{R}^{L_z}$ and $\mathbf{v}_{\phi_{\mathbf{z}}, t} \in \mathbb{R}^{L_z}_{+}$ depend on $\{\mathbf{s}_{1:T}, \mathbf{w}\}$.

\subsection{Implementation}\label{sec:implementation}

To implement the above equations, LigHT-DVAE is built on the original Transformer architecture~\cite{10.5555/3295222.3295349}, with several notable differences, all visible in Fig.~\ref{fig:model_structure}: (a) There are two decoders working in parallel: one to compute the parameters of \eqref{eq:gen-z} and one to compute the parameters of \eqref{eq:gen-s}; (b) The output of the encoder and decoders are distribution parameters (in place of ``deterministic'' data representations in the conventional use of the Transformer); (c) The output target of the (second) decoder is the input data $\mathbf{s}_{1:T}$ (as opposed to a different sequence in regression/translation problems); (d) A sequence-wise variable $\mathbf{w}$ is introduced; (e) In the cross-attention module of the decoders, the role of $\mathbf{z}_{1:T}$ (encoder output) and $\mathbf{s}_{1:T}$ (decoder output) as source information for the query and key/value, respectively, is inverted compared to the conventional Transformer. Points (a), (b) and (c) are inherited from the DVAE framework. Point (d) is a model design choice (see Section~\ref{subsec:generative_model}). Point (e) will be justified later.  

In a few more details, encoding actually starts with the encoding
of the sequence-wise latent variable $\mathbf{w}$. An RNN is used to compute $\boldsymbol{\mu}_{\phi_{\mathbf{w}}}$ and $\mathbf{v}_{\phi_{\mathbf{w}}}$ from $\mathbf{s}_{1:T}$ (not represented in Fig.~\ref{fig:model_structure}). We only retain the RNN output at the last time index $T$, since it contains all information on the complete sequence $\mathbf{s}_{1:T}$. The sampled value of $\mathbf{w}$ is then replicated $T$ times and concatenated to $\mathbf{s}_{1:T}$ to produce the input of the Transformer encoder. This latter follows the line of the original Transformer \cite{10.5555/3295222.3295349} (see Fig.~\ref{fig:model_structure}): input embedding followed by positional encoding, followed by a Multi-Head Attention (MHA) module, normalization layer (NL) with skip connection with the MHA input, feed-forward (FF) layer and another NL with skip connection with the FF layer input. Finally, a linear layer produces the parameters of  \eqref{eq:inf-z}, that is $\boldsymbol{\mu}_{\phi_{\mathbf{z}},t} $ and the logarithm of $\mathbf{v}_{\phi_{\mathbf{z}}, t}$.

 
The two decoders also follows the general line of the original Transformer \cite{10.5555/3295222.3295349} (see Fig.~\ref{fig:model_structure}; here we do not detail the different modules as we did for the encoder, because of room limitation; see \cite{10.5555/3295222.3295349} for details). The sampled value of $\mathbf{z}_{1:T}$ is concatenated with the replicated sampled value of $\mathbf{w}$ to produce the ``main'' input, which is the source information for the query of the second MHA module. In parallel, $\mathbf{s}_{1:T}$ is used as the second input, i.e. the source information for the key and value of the second MHA module. A causal mask in the MHA module ensures that the dependency at time $t$ is on $\mathbf{s}_{1:t-1}$.   

The difference between the two decoders essentially consists of different masks in the first MHA module. For the first decoder (generating $\boldsymbol{\mu}_{\theta_{\mathbf{z}},t}$ and $\mathbf{v}_{\theta_{\mathbf{z}}, t}$), this mask is set to limit the input to $\mathbf{z}_{1:t-1}$, whereas for the second decoder (generating $\mathbf{v}_{\theta_{\mathbf{s}}, t}$), it is set to limit the input to $\mathbf{z}_{1:t}$. In short, the masks in the MHA modules are adapted to implement the causal dependencies defined in \eqref{generative-joint-distribution-hit-dvae}.

As stated above, a main difference of LigHT-DVAE w.r.t.\ the original Transformer is that the queries to decode $\mathbf{s}_t$ are computed from $\mathbf{z}_{1:t}$ and not $\mathbf{s}_{1:t-1}$. In fact, the residual connections in the decoders will make the output rely more on the information from the queries. If we use the previous $\mathbf{s}$ vectors to compute the queries, this will carry too much information and learning does not generalize well. However, while this phenomena is observed on $\mathbf{s}$, we have no reason to believe that it should reproduce on $\mathbf{z}$. Unlike HiT-DVAE, we propose to query both generation processes with $\mathbf{z}$, thus allowing the two decoders to share the parameters.

Similar to other DVAE models \cite{MAL-089}, LigHT-DVAE is trained by maximizing the following evidence lower bound (ELBO):
\begin{align}\label{eq:ELBO-HiT-DVAE}
    \mathcal{L}&(\theta, \phi; \mathbf{s}_{1:T}) =  - D_{\text{KL}}(q_{\phi_{\mathbf{w}}}(\mathbf{w}|\mathbf{s}_{1:T}) \parallel p_{\theta_{\mathbf{w}}}(\mathbf{w}))\nonumber\\ &-\sum_{t=1}^{T}\mathbb{E}_{q_{\phi_{\mathbf{z}}}q_{\phi_{\mathbf{w}}}}\big[ d_{\text{IS}}(|\mathbf{s}_{t}|^2, \mathbf{v}_{\theta_{\mathbf{s}},t}) \nonumber\\
    & + D_{\text{KL}}(q_{\phi_{\mathbf{z}}}(\mathbf{z}_t|\mathbf{s}_{1:T}, \mathbf{w}) \parallel p_{\theta_{\mathbf{z}}}(\mathbf{z}_t|\mathbf{s}_{1:t-1}, \mathbf{z}_{1:t-1}, \mathbf{w})) \big]
\end{align}
where $d_{\text{IS}}(\cdot,\cdot) $ is the Itakura-Saito divergence \cite{6797100}, $D_{\text{KL}}(\cdot||\cdot)$ is the Kullback–Leibler divergence (KLD), and square is element-wise.

\section{Experiments}
\label{sec:experiments} 
\subsection{Datasets and pre-processing}
The experiments are conducted on two datasets: the Wall Street Journal (WSJ0) dataset \cite{WSJ0} and the Voice Bank (VB) corpus \cite{valentini2016speech}. The WSJ0 dataset is composed of 16-kHz speech signals, with three subsets: \textit{si\_tr\_s}, \textit{si\_dt\_05} and \textit{si\_et\_05}, used for model training, validation and test, and containing 12,765, 1,026 and 651 utterances respectively. The VB dataset contains a training set with 11,572 utterances performed by 28 speakers and a test set with 824 utterances performed by 2 speakers. 
We follow \cite{fu2021metricganU} to choose two speakers (p226 and p287) from the training set for validation, which contains 770 utterances, and use the leftover 26 speakers for training.

In all experiments, the raw audio signals are pre-processed in the following way. 
First, the silence at the beginning and the end of the signals are cropped by using a voice activity detection threshold of 30 dB. Then, the waveform signals are normalized so that their maximum absolute value is one. The STFT is computed with a 64-ms sine window (1024 samples) and a 75\%-overlap (256 samples hop length), resulting in a sequence of 513-dimensional discrete Fourier coefficients (for positive frequencies). Finally, the power magnitude of the STFT coefficients is computed. We set the sequence length of each STFT spectrogram to $T=150$ (corresponding to speech segments of 2.4s) for WSJ0 and $T=100$ (corresponding to speech segments of 1.6s) for VB. 
At test time, the model is evaluated on the complete test utterances which can be of variable length.

\subsection{Implementation details and training settings}\label{subsec:implementation}

The Transformer-based encoder and decoders are composed of 4 identical layers as described in Section~\ref{sec:implementation}. All input vectors are embedded into vectors of dimension $d_{\text{model}}=256$, which is the size for all MHA blocks. We apply single-head attention because we found that using multi-head attention decreases the performance in our experiments. All feed-forward blocks in the Transformer layers consist of two dense layers with size 1024 and 256. The latent dimension $L_{z}$ and $L_{w}$ are set to $16$ and $32$, respectively.


The training is made with the AdamW \cite{loshchilov2018decoupled} optimizer, which is a variant of Adam \cite{DBLP:journals/corr/KingmaB14} with decoupled weight decay. The parameters of the optimizer are $\beta_1=0.9, \beta_2=0.99, \epsilon=10^{-9}$ and $\texttt{weight\_decay}=10^{-5}$. We varied the learning rate during training by firstly increasing it linearly for the warm-up training stage (for 5k iterations) and then decreasing it using a cosine annealing scheduler \cite{loshchilov2017sgdr} (for 20k iterations). The maximum and minimum values of the learning rate are $5 \times 10^{-5}$ and $10^{-8}$ respectively. $\beta$ factors are multiplied to the KLD terms in \eqref{eq:ELBO-HiT-DVAE} to increase latent space expressivity, with $\beta_w = \beta_z = 10^{-2}$.

\subsection{Baselines and evaluation metrics}
We compare HiT-DVAE and LigHT-DVAE with the basic VAE and three other DVAE models: The Deep Kalman Filter (DKF) \cite{https://doi.org/10.48550/arxiv.1511.05121}, the Recurrent Variational AutoEncoder (RVAE) \cite{9053164}, and the Stochastic Recurrent Neural Network (SRNN) \cite{10.5555/3157096.3157343}. As HiT/LigHT-DVAE, SRNN is an AR model that uses $\mathbf{s}_{1:t-1}$ to generate $\mathbf{s}_t$. DKF and RVAE are not AR. DKF generate $\mathbf{s}_t$ from $\mathbf{z}_t$, and RVAE does it from $\mathbf{z}_{1:t}$ (we use the causal version of RVAE; see \cite{9053164}). As mentioned in the introduction, an AR model can be trained either in \textit{teacher-forcing} mode or in \textit{scheduled-sampling} mode. In our experiments, SRNN is trained in both modes, while HiT/LigHT-DVAE are only trained in the teacher-forcing mode. Except for that, all models are trained with the same settings described in Section~\ref{subsec:implementation}. We evaluate the average speech analysis-resynthesis performance on the test set (in generation mode for AR models) using four evaluation metrics: The root mean squared error (RMSE), the scale-invariant signal-to-distortion ratio (SI-SDR) \cite{8683855} in dB, the perceptual evaluation of speech quality (PESQ) score \cite{941023} (in $[-0.5, 4.5]$), and the extended short-time objective intelligibility (ESTOI) score \cite{5713237} (in $[0,1]$). We also evaluate the average generation performance via the Fréchet Deep Speech Distance (FDSD) proposed in \cite{Binkowski2020High}. FDSD is a quantitative metric for speech signals generative models. It relies on speech features extracted with the pre-trained speech recognition model DeepSpeech2 \cite{10.5555/3045390.3045410}.

\begin{table}[t]
\centering
\resizebox{\linewidth}{!}{
 \begin{tabular}{cccccc} 
\toprule
Dataset & Model & RMSE $\downarrow$ & SI-SDR $\uparrow$ & PESQ $\uparrow$ & ESTOI $\uparrow$ \\
\midrule
\multirow{8}*{WSJ0} & VAE & 0.040 & 7.4 & 3.28 & 0.88 \\
~ & DKF & 0.037 & 8.3 & 3.51 & \textbf{0.91} \\
~ & RVAE & 0.034 & 8.9 & 3.53 & \textbf{0.91} \\
~ & SRNN (SS) & 0.036 & 8.7 & \textbf{3.57} & \textbf{0.91} \\
~ & SRNN (TF) & 0.061 & 2.6 & 2.53 & 0.76 \\
~ & HiT-DVAE & 0.031 & 10.0 & 3.52 & \textbf{0.91} \\
~ & LigHT-DVAE & \textbf{0.030} & \textbf{10.1} & 3.55 & \textbf{0.91} \\
\midrule
\multirow{7}*{VB} & VAE & 0.052 & 8.4 & 3.24 & 0.89 \\
~ & DKF & 0.048 & 9.3 & 3.44 & 0.91 \\
~ & RVAE & 0.050 & 8.9 & 3.39 & 0.90 \\
~ & SRNN (SS) & 0.044 & 10.1 & 3.42 & 0.91 \\
~ & SRNN (TF) & 0.102 & -0.1 & 2.15 & 0.75 \\
~ & HiT-DVAE & 0.039 & 11.4 & \textbf{3.60} & \textbf{0.93} \\
~ & LigHT-DVAE & \textbf{0.038} & \textbf{11.6} & 3.58 & \textbf{0.93} \\
\bottomrule
\end{tabular}
}
\caption{Speech spectrograms analysis-resynthesis results.} 
\label{table:analysis-resynthesis-results}
\end{table}

\subsection{Speech spectrograms analysis-resynthesis results}

The speech analysis-resynthesis results on both the WSJ0 dataset and the VB dataset are reported in Table~\ref{table:analysis-resynthesis-results}. 
On the WSJ0 dataset, HiT-DVAE and LigHT-DVAE outperform the other DVAE models for all metrics, except for PESQ (SRNN (SS) is slightly better) and ESTOI (the five models are on par).
On the VB dataset, HiT/LigHT-DVAE  outperform all other DVAE models on all metrics. As AR models, even if HiT/LigHT-DVAE are trained in teacher-forcing mode, they still keep very robust performance when tested in generation mode. In contrast, SRNN (TF) trained in teacher-forcing mode leads to a notable drop of performance compared to SRNN (SS) trained in scheduled-sampling mode. Finally, the experimental results show that sharing the parameters of the decoders for $\mathbf{s}$ and $\mathbf{z}$ in LigHT-DVAE slightly increases the performance compared to HiT-DVAE. At the same time, it reduces the number of model parameters from 21.75 M to 17.46 M.

\subsection{Ablation studies on model structures}

To try to explain why HiT/LigHT-DVAE are robust to the teacher-forcing training mode, we performed several ablation studies on the models structure on the VB dataset. As explained in Section~\ref{sec:implementation}, different to the original Transformer decoder, the queries to decode $\mathbf{s}_t$ in HiT/LigHT-DVAE are computed from $\mathbf{z}_{1:t}$ instead of $\mathbf{s}_{1:t-1}$. In a first ablation experiment, we invert the role of the queries and the keys/values when decoding $\mathbf{s}_{t}$, i.e., we swap the decoder inputs $\mathbf{z}_{1:t}$ and $\mathbf{s}_{1:t-1}$. the resulting models are denoted with ``Inv-s''. The models are always trained in the teacher-forcing mode. Table \ref{table:ablation-hit-dvae} reports the evaluation results when using generated or ground-truth $\mathbf{s}_{1:t-1}$ to decode $\mathbf{s}_{t}$, in the ``GEN'' and ``GT'' rows respectively.

\begin{table}[!t]
\centering
\resizebox{\linewidth}{!}{
\begin{tabular}{cccccc} 
\toprule
Test $\mathbf{s}_{1:t-1}$ & Model & RMSE $\downarrow$ & SI-SDR $\uparrow$ & PESQ $\uparrow$ & ESTOI $\uparrow$ \\
\midrule
\multirow{6}*{GEN} & HiT-DVAE & 0.039 & 11.4 & 3.60 & 0.93 \\
~ & HiT-DVAE-Inv-s & 0.079 & 3.8 & 2.61 & 0.75 \\
~ & HiT-DVAE-Inv-s-NR & 0.067 & 5.8 & 2.68 & 0.78 \\
~ & LigHT-DVAE & 0.038 & 11.6 & 3.58 & 0.93 \\
~ & LigHT-DVAE-Inv-s & 0.079 & 3.9 & 2.58 & 0.75 \\
~ & LigHT-DVAE-Inv-s-NR & 0.068 & 5.7 & 2.63 & 0.78 \\
\midrule
\multirow{6}*{GT} & HiT-DVAE & 0.038 & 11.5 & 3.60 & 0.93 \\
~ & HiT-DVAE-Inv-s & 0.038 & 11.4 & 3.32 & 0.90 \\
~ & HiT-DVAE-Inv-s-NR & 0.067 & 5.8 & 2.68 & 0.78 \\
~ & LigHT-DVAE & 0.038 & 11.7 & 3.59 & 0.93 \\
~ & LigHT-DVAE-Inv-s & 0.040 & 10.9 & 3.29 & 0.89 \\
~ & LigHT-DVAE-Inv-s-NR & 0.068 & 5.7 & 2.63 & 0.78 \\
\bottomrule
\end{tabular}
}
\caption{Speech spectrograms analysis-resynthesis results: Ablation studies on the HiT/LigHT-DVAE models structure.} 
\label{table:ablation-hit-dvae}
\end{table}

Although the Inv-s models achieve very good results when trained and evaluated in TF mode, their performance notably drops when evaluating in GEN mode (e.g., from 11.4 to 3.8 dB SI-SDR for HiT-DVAE-Inv-s). 
We believe that this lack of robustness to a mismatch between the training and test modes is mainly caused by the residual connections on the queries in the original Transformer decoder architecture (which are highlighted in bold in Figure \ref{fig:model_structure}). Indeed, in the Inv-s models, $\mathbf{s}_{1:t-1}$ is used to compute the queries and the residual connections make the prediction of $\mathbf{s}_t$ directly relying on the previous vectors $\mathbf{s}_{1:t-1}$. This is not the case in the HiT/LigHT-DVAE models where $\mathbf{z}_{1:t-1}$ is used to compute the queries from which residual connections start. This forces the prediction of $\mathbf{s}_t$ to rely more on $\mathbf{z}_{1:t-1}$ than $\mathbf{s}_{1:t-1}$, which explains the better generalization of HiT/LigHT-DVAE compared to HiT/LigHT-DVAE-Inv-s. This structural aspect is very important in the present context of speech modeling, because adjacent speech spectrogram frames are much more correlated than discrete tokens in natural language processing, the original application domain of Transformers. 
To confirm this interpretation, we further removed the residual connections in the decoder of $\mathbf{s}$ in the Inv-s models (this configuration is referred to as Inv-s-NR models). As a result, the performance slightly increased compared to the Inv-s models in the GEN test mode, but it remains much below that of HiT/LigHT-DVAE. Also, the performance of the Inv-s-NR models is the same when evaluated in GEN mode and GT mode, confirming that the generalization problem of the Inv-s models was due to the use of $\mathbf{s}_{1:t-1}$ to compute the queries from which residual connections start. Overall, this ablation study experimentally showed the importance of the architectural choices made in the HiT/LigHT-DVAE models for the modeling of highly correlated sequences of continuous data such as speech spectrograms.

\subsection{Speech spectrograms generation results} In addition to the above speech analysis-resynthesis experiments, we also evaluated the models performance in speech spectrogram generation via FDSD scores. To compute these scores, we generated $10,\!240$ $1.6$-s speech signal samples from each model. The DVAE models generate power spectrograms, from which we deduced magnitude spectrograms. We then used the Griffin-Lim  algorithm \cite{1164317} to reconstruct the phase spectrograms and generate the waveform signals. We used 1.6-s utterances from the VB training set as the reference set to compute the FDSD scores. 
Finally, in order to have a reference FDSD score, we also computed the FDSD on the VB test set, with exact original phase or phase reconstructed by the Griffin-Lim algorithm.
As shown in Table~\ref{table:generation}, HiT-DVAE outperforms the other DVAE models on this generation task. This is a new result, since HiT-DVAE was never used before for speech modeling. It can also be seen that the AR DVAE models (HiT/LigHT-DVAE and SRNN) generally have a better generation performance than the non-AR DVAE models.

\begin{table}[!t]
\centering
\resizebox{.58\linewidth}{!}{
 \begin{tabular}{cc} 
\toprule
Model & FDSD $\downarrow$ \\
\midrule
VAE & 70.92 ± 0.44 \\
DKF & 32.78 ± 0.28 \\
RVAE & 45.75 ± 0.11 \\
SRNN (SS) & 25.28 ± 0.19 \\
SRNN (TF) & 25.53 ± 0.13 \\
HiT-DVAE & \textbf{22.50 ± 0.26} \\
LigHT-DVAE & 29.22 ± 0.26 \\
\midrule
VB Test (exact phase) & 4.11 ± 0.14 \\
VB Test (Griffin-Lim) & 4.11 ± 0.15 \\
\bottomrule
\end{tabular}
}
\caption{Speech spectrograms generation results.}
\label{table:generation}
\end{table}

\section{Conclusion}
\label{sec:conclusions}
In this paper, we applied the HiT-DVAE model of \cite{https://doi.org/10.48550/arxiv.2204.01565} and its new variant LigHT-DVAE on speech signals and have demonstrated their strong modeling capacity through experiments on speech spectrograms analysis-resynthesis and generation. Although HiT/LigHT-DVAE are autoregressive models, they have shown to be very robust to the mismatch between the teacher-forcing mode at training time and the generation mode at test time, as opposed to previous AR DVAE models such as SRNN. This enables using a simpler training procedure. Compared to HiT-DVAE, the new proposed LigHT-DVAE model exhibits competitive performance with about $20\%$ less parameters. 
We believe that HiT/LigHT-DVAE have a great potential in unsupervised speech representation learning and dynamical modeling, which can benefit downstream speech processing tasks such as speech enhancement, speech separation and speech inpainting.



\vfill
\pagebreak

\bibliographystyle{IEEEbib}
\bibliography{refs}

\end{document}